\documentstyle[12pt]{article}

\def\+{{+\!\!\!+}}

\def\d{\partial}

\def\l{\lambda}

\def\s{\sigma}

\def\t{\tilde}

\def\pmb#1{\setbox0=\hbox{#1}%
\kern.0em\copy0\kern-\wd0
\kern-.04em\copy0\kern-\wd0
\kern.08em\copy0\kern-\wd0
\kern-.04em\raise.0433em\box0 }         

\newcommand{\nc}{\newcommand}
\nc{\beq}{\begin{equation}}
\nc{\eeq}[1]{\label{#1}\end{equation}}
\nc{\ber}{\begin{eqnarray}}
\nc{\eer}[1]{\label{#1}\end{eqnarray}}
\nc{\pek}[1]{\cite{#1}}
\nc{\enr}[1]{(\ref{#1})}
\nc{\kal}[1]{{\cal{#1}}}
\nc{\dott}{\;\cdot\;}

\begin{document}
\newcommand{\inv}[1]{{#1}^{-1}} 
\renewcommand{\theequation}{\thesection.\arabic{equation}}
\newcommand{\be}{\begin{equation}}
\newcommand{\ee}{\end{equation}}
\newcommand{\bea}{\begin{eqnarray}}
\newcommand{\eea}{\end{eqnarray}}
\newcommand{\re}[1]{(\ref{#1})}
\newcommand{\qv}{\quad ,}
\newcommand{\qp}{\quad .}


\pagestyle{empty}
\begin{center}

                                \hfill    USITP-00-07\\
                                \hfill    hep-th/0005202\\

\vskip .3in \noindent

\vskip .1in

{\large \bf {Bosonic Open Strings in a Tachyonic Background Field}}
\vskip .2in

{\bf Tasneem Zehra Husain}\footnote{tasneem@physto.se}
 and {\bf Maxim Zabzine}\footnote{zabzin@physto.se}  \\

\vskip .15in

\vskip .15in

\vskip .15in

 {\em Institute of Theoretical Physics,
University of Stockholm \\
Box 6730,
S-113 85 Stockholm SWEDEN} \\
\bigskip

\vskip .15in

\vskip .1in
\end{center}
\vskip .4in
\begin{center} {\bf Abstract}
\end{center}
\begin{quotation}\noindent
 We study a bosonic open string coupled to a tachyonic
 background field $T[X]$ and find that the tachyon field can 
 effectively be replaced by a configuration of D-branes placed at either 
 the zeros or the critical points of $T[X]$, depending on our choice of 
 boundary conditions. This dual picture of the open string tachyon 
 is explored in detail for the explicit case when the tachyonic field
 is quadratic.    
\end{quotation}
\vskip 1.2in
\begin{center}
{\bf Talk presented at the Nordic Network On String Theory}\\ 
{\bf held at the Niels Bohr Institute, Copenhagen in May 2000.} 
\end{center}
\vfill
\eject

\pagenumbering{arabic}
\clearpage
\setcounter{page}{1}

\section{Introduction and motivation}

 Recent interest in the construction of non-BPS branes 
 \cite{Sen:1998sm} has led to the
 investigation of relationships between Dp-branes in which the 
 tachyonic field figures prominently. 
 In particular, it has been established that a D-brane can be thought of 
 as the solitonic solution of a tachyonic field living on a higher
 dimensional brane-antibrane system \footnote{The analysis can be 
 extended to a tachyonic field depending on more than one 
 spatial directions\cite{Majumder:2000tt}.}. Sen showed that a 
 $D(p-1)$-brane can be identified with 
 the tachyonic kink on a $Dp$-$\bar{D p}$ system, or alternatively,
 with the tachyonic lump on a single Dp-brane \cite{Sen:1999mh}; the latter
 picture is supported by calculations in string bosonic field theory  
 \cite{Harvey:2000tv}. Subsequently, in \cite{Harvey:2000na}, these lumps were 
 interpreted in terms of relevant boundary perturbations of the 
 world-sheet CFT. To illustrate some of the salient features of this 
 description, we consider a model in which a bosonic open string couples to 
 a tachyonic background field $T[X]$ (in flat Minkowski space-time), and look
 for the consistent open string configurations which may exist.

 The paper is organized as follows: We start in Section $2$ with an
 arbitrary tachyonic background field and a diffeomorphism
 invariant world-sheet action, analyzing all possible situations in which 
 a string interpretation can be obtained. World-sheet 
 parity plays a critial role in the analysis.  We find that
 locally, D-branes can be associated with either tachyonic kinks or lumps. 
 To make these ideas explicit, we focus on a 
 quadratic tachyonic profile in Section $3$. This is an instructive example 
 as it illustrates the main features under discussion, yet is a model 
 which can be solved explicity. In Section $4$, we synthesize the results 
 of our analysis. The Appendix contains a brief outline of how one might 
 deal with a more general profile. 

\section{A General Analysis}

 Consider a bosonic open string moving in a 
 flat background with metric $\eta_{\mu\nu}=(-1,1,...,1)$ and coupling   
 to a tachyonic profile $T[X]$. The diffeomorphism invariant action 
 is:
\be
\label{action}
 S= -\frac{1}{4\pi\alpha'} \int\limits_{\Sigma}
  d^2\sigma\,\,\sqrt{-h} h^{\alpha\beta}
 \d_{\alpha} X^\mu \d_\beta X^\nu \eta_{\mu\nu} + 
 \int\limits_{\d \Sigma} d\tau\,\sqrt{-h_{00}} T[X].
\ee
 This model is Weyl invariant in the bulk but not at the boundary
\be
\label{a1}
 h_{\alpha\beta} \rightarrow \Lambda(\sigma) h_{\alpha\beta},
\,\,\,\,\,\,\,\,\,\,\,\,\,\,\,\Lambda|_{\partial \Sigma}=0.
\ee
 Despite the fact that the boundary term does not preserve Weyl invariance, 
 we still have enough symmetries to impose conformal gauge, since the 
 metric dependence 
 in the boundary term can be completely gauged away using only diffeomorphism 
 invariance\footnote {In general this is a tricky question. Fixing a boundary
 parameterization is equivalent to imposing boundary conditions on the bulk 
 diffeomorphisms.}. In conformal gauge the constraints are given by:
\ber
 T_{00} = \frac{1}{2}(\dot{X}^2 + X'^2) + 2\pi\alpha' [\delta(\sigma) T[X]
 + \delta(\pi-\sigma)T[X]] = 0,\label{a3}\\
 T_{01}=T_{10}= \dot{X}\cdot X' = 0,
\,\,\,\,\,\,\,\,\,\,\,\,\,\,\,\,\,\,\,\,
T_{11} = \frac{1}{2}(\dot{X}^2 + X'^2) = 0\label{a4}.
\eer{TT}
 Consistency of the model forces us to require that along with the 
 standard Virasoro conditions $(\dot{X}\pm X')^2=0$, the condition 
 $[\delta(\sigma) T[X] + \delta(\pi-\sigma)T[X]] =0$
 also be fulfilled classically, restoring Weyl invariance in the 
 original action. Re-writing this as $T[X]|_{\s=0,\pi}=0$,
 we see that it is equivalent to imposing Dirichlet boundary 
 conditions on $X^i$ in such a way that string end-points are confined 
 to zeros of $T[X]$.

 Since the tachyonic field does not couple to the bulk, 
 the equation of motion is unchanged and we have, as usual,
 $\d_\alpha \d^\alpha X^\mu =0$. 
 However, the boundary term does change and 
we can now impose either Dirichlet or "modified Neumann" boundary conditions
\be
\label{mmmm}
\delta X^\mu|_{0,\pi} =0,\,\,\,\,\,\,\,\,\,\,\,\, 
\,\,\,\,\,\,\,\,\,\,\,\,
(\eta_{\mu\nu}X'^\nu - 2\pi\alpha' \d_\mu T[X])|_{\s=0,\pi} =0. 
\ee
Consider a tachyonic field which depends on coordinates $X^i$
($i=p+1,...,d-1$). For the sake of simplicity, we impose Neumann 
boundary conditions on directions $X^n$, ($n=0,1,...,p$) transverse to 
the profile. 
Along $X^i$ however, we have Dirichlet boundary conditions and the  
following mode expansion
\be
 X^i(\tau,\sigma)= q^i + b_0^i\s + \sum\limits_{n\neq 0} 
 \frac{e^{-in\tau}}{n}  b_n^\mu \sin n\sigma
\ee 
The positions of the Dp-branes are given by 
$X^i|_0 = q^i$ and $X^i|_{\pi} = q^i + b^i_0 \pi$, where $q^i$ and $b_0^i$
satisfy $T[q]=0 = T[q + b_0\pi]=0.$ Hence, zeros of a tachyonic profile can be 
 thought of as positions of Dp-branes which span the p directions 
 transverse to the tachyon field. 
 This is rather remarkable statement as it enables us
 to 'replace' a tachyon field with a number of D-p branes in the appropriate
 positions, an interpretation supported by results obtained in the closed 
 string channel. 

 To illustrate this point we consider a tachyonic field depending 
 on only one spatial direction, and proceed to construct the bosonic 
 boundary state $|B_X\rangle$.
 Open and closed string channels are related through the following 
 correspondence:
\be
\label{corrbos}
\left \{ \begin{array}{l}
  \d_\sigma X^n |_{\s=0}=0\\
  T[X]|_{\s=0} = 0
\end{array} \right . ,\,\,\,\,\,\,\,\,\,\,\Rightarrow \,\,\,\,\,\,\,\,\,\,
\left \{ \begin{array}{l}
\d_\tau X^n|_{\tau=0}|B_X\rangle =0 \\
 T[X]|_{\tau =0}|B_X\rangle = 0.
\end{array} \right .
\ee
 One can solve these conditions and construct the boundary state
\be
\label{BX}
 |B_X \rangle = N \sum\limits_{q_n} \frac{1}{T'[q_n]}
 \delta(q - q_n) \left( \prod\limits_{n = 1}^{\infty}
e^{- \frac {1}{n} a_{-n}S\t{a}_{n}} \right)
 |0\rangle|_{a} |0\rangle_{\t{a}} |p = 0 \rangle.
\ee
 Here, N is a normalization constant 
 (given in \cite{DiVecchia:1999rh}, eq.(6.180)) and we have 
 used the following identity
\be
\label{deltaf}
 \delta(f(x)) = \sum\limits_n \frac{\delta(x-x_n)}{|f'(x_n)|}
\ee
 where $f(x)\in C^{\infty}$ and $x_n$ are simple zeros of $f(x)$. Thus at
 every simple zero of $T[X]$ we can place one D p-brane.

 While the positions of the branes are dictated by zeros of the field, the 
 number of D p-branes at a particular position is determined by the order 
 of the zero. As a simple illustration of the above statement, consider 
 $T[X] \sim  (X-X_0 + \epsilon)(X-X_0 - \epsilon)$. 
There is now one D p-brane each localised at $(X_0 - \epsilon)$ and 
 $(X_0 + \epsilon)$, with a U(1) gauge theory on its world-volume.
 As $\epsilon \rightarrow 0$ the Dp branes approach each other, meeting 
 at $X_0$, so we end up with 2 coincident branes and a $U(2)$ 
 world-volume gauge theory. Reversing this logic, we can seperate 2 coincident
 branes and thus break gauge symmetry from $U(2)$ to $U(1) \times U(1)$. 
 The masses of the Higgs and W-bosons will be determined by the form 
 of the tachyon field, or more precisely, the distance between its zeros. 

 Now consider the modified Neumann boundary conditions 
 in (\ref{mmmm}). 
  In general they break conformal invariance, ruling out the possibility 
 of a string interpretation. 
 However one can follow the logic proposed by Harvey et.al 
 \cite{Harvey:2000na} and think of the boundary interaction as a 
 relevant operator which induces a world-sheet renormalization group flow
 such that the string picture is recovered at the  
 IR and UV fixed points.

 We apply this procedure to our theory, defining 
 coordinates $\phi^\mu = X^\mu/\sqrt{2\pi\alpha'}$ which enable us 
 to canonically normalize the scalar in two dimensions. 
 In conformal gauge, boundary interactions have the following form
\be
\label{biV}
 \pm \lambda \int d\tau V(\phi)|_{0} \pm \lambda \int d\tau V(\phi)|_{\pi}
\ee
 where $\lambda$ is a dimensionful coupling constant\footnote{Generically there
 may be many coupling constants, however for the time being we restrict
 ourselves to the case where there is only one} and $V(\phi)$
 is a dimensionless potential given by  
 $\lambda V(\phi) = T(\sqrt{2\pi\alpha'}\phi)$. A priori there are no
 restrictions on the signs in front of boundary
 couplings in (\ref{biV}) so we have four possibilities 
 : $(+,+)$, $(+,-)$, $(-,+)$ and $(-,-)$. Each choice corresponds to a 
  different action and hence a different theory. We use these signs to 
 label the theories. Boundary conditions for theories $(+,+)$ and $(+,-)$ are
\be
\label{potdim}  
 (\eta_{ij}\phi'^j + \lambda \partial_i V(\phi))|_{0} =0,
\,\,\,\,\,\,\,\,\,\,\,\,
(\eta_{ij}\phi'^j \mp \lambda \partial_i V(\phi))|_{\pi} =0 
\ee
respectively, whereas for theories $(-,-)$ and $(-,+)$ we find
\be
\label{potdim2}  
 (\eta_{ij}\phi'^j - \lambda \partial_i V(\phi))|_{0} =0,
\,\,\,\,\,\,\,\,\,\,\,\,
(\eta_{ij}\phi'^j \pm \lambda \partial_i V(\phi))|_{\pi} =0 
\ee
 Theories $(+,+)$ and $(-,-)$ are invariant under worldsheet parity 
 $\Omega$ ($\sigma \rightarrow \pi -\s$) while theories $(+,-)$
 and $(-,+)$ are interchanged. Since $\Omega$ is a global 
 symmetry of open string theory and is expected to be preserved, 
 we restrict future analysis to the parity invariant theories only. 
 
 The parameter $\lambda$ has dimension of mass, so one can naively consider 
 $\lambda V(\phi)$ as a relevant boundary 
 perturbation\footnote{In general one should be careful in determining whether 
 an operator is relevant or not; a relevant operator should have conformal 
 weight less than one.}. Thus, in each theory, we have some 
 massless scalars in the bulk and a relevant boundary perturbation which 
 breaks conformal invariance explicitly. At the fixed points we find 
 the following boundary conditions
 \be
\label{fpbc}
\phi'^i|_{\s=0,\pi} =0 \,\, at\,\,\lambda=0,
\,\,\,\,\,\,\,\,\,\,\,\,\,\,\,\,\,\,
\partial_i V(\phi)|_{\s=0,\pi} =0 \,\, at \,\,\lambda=\infty.  
\ee  
 Again, we can interpret the latter as Dirichlet boundary conditions. So 
 the relevant boundary perturbation induces a RG flow from the UV fixed point 
 with Neumann boundary conditions, to the IR fixed point where we have D 
 p-branes localized at the critical points of $V$; these could be 
 maxima or minima depending on the theories under consideration. In the 
 Dirichlet case, we require that $V(\phi) = 0$ at the endpoints 
 of the string, as this is the only way to restore Weyl invariance in 
 the theory. For the modified Neumann case this condition is no longer needed,
 since Weyl invariance is restored anyway at the IR and UV fixed points. 

To summarize then, we find that Dirichlet conditions along $X^i$ lead 
to Dp-branes localized at zeros of $T[X]$ and resembling tachyonic kinks, 
whereas modified Neumann boundary conditions lead to Dp-branes which are
localized at the maxima/minima of $T[X]$ and are reminescent of tachyonic 
lumps. 

Since Dp-branes can be localized either at minima or at maxima of 
$V(\phi)$ in the IR, it is important that perturbation theory be
well defined for both cases. 
Let us argue that this is true for a boundary potential. For the sake 
of simplicity we drop space-time indices. The Minkowski world-sheet 
action for the $(+,+)$ theory can then be expressed as follows:
\be
\label{Mwsa}
 S= \frac{1}{2} \int d^2\s [(\dot{\phi})^2 - (\phi')^2]
 + \l \int d\tau V(\phi)|_0 + \l \int d\tau V(\phi)|_\pi,
\ee  
Inorder to use the 
path integral formalism we make the Wick rotation $\tau = i t$.
ssuming that Taylor expansion is valid around the extrema 
$\phi_0$ and $\phi_1$ where the ends of the string are localized, we have
\be
\label{teVp}
 V(\phi) = \sum\limits_{n=0}^{\infty} \frac{1}{n!} V^{(n)}(\phi_0) 
 (\phi-\phi_0)^n \;
\; {\rm for}\; \s = 0 \;\; {\rm and} \;\;
 V(\phi) = \sum\limits_{n=0}^{\infty} \frac{1}{n!} V^{(n)}(\phi_1) 
 (\phi-\phi_1)^n
\;\; {\rm for} \; \s = \pi.
\ee
The Euclidean action can then be written as
 $$
- i S = - \frac{1}{2} \int d^2\s\,\phi \Box \phi  +  \l V(\phi_0)\beta +
\l V(\phi_1)\beta  
- \frac{\l}{2} \sum\limits_{n=1}^{\infty} \frac{(n-2)}{n!} V^{(n)}(\phi_0) 
 \int dt (\phi-\phi_0)^n|_0$$ 
\be
\label{AETE}
- \frac{\l}{2}\phi_0 \sum\limits_{n=1}^{\infty} 
 \frac{1}{(n-1)!} V^{(n)}(\phi_0) 
 \int dt (\phi-\phi_0)^{n-1}|_0 - ...
\ee
 where $...$ denotes a similar contribution from the boundary at $\pi$. 
 To avoid IR singularities we introduce finite temperature $\beta = 1/T$ 
 and consider the theory on an annulus. 
 
 Dropping $V(\phi_0)$ and $V(\phi_1)$ and taking into account that 
 $V'(\phi_0)=0$ and $V'(\phi_1)=0$ we find that second derivatives appear
 in the following terms
\be
- \frac{\l\beta}{2} [\phi_0^2 V^{(2)}(\phi_0) + \phi_1^2 V^{(2)}(\phi_1)]
 +  \frac{\l}{2}[\phi_0 V^{(2)}(\phi_0) \int dt\,\phi|_0 + 
\phi_1 V^{(2)}(\phi_1) \int dt\,\phi|_\pi] .
\ee 

 If both ends of the 'string' lie at the same critical point, $\phi_0=\phi_1$,
 the field $\phi$ can be shifted in such way that
 $V^{(2)}$ term drops out completely (contrary to what happens with a bulk 
 potential) so locally there is no way to distinguish between maxima and 
 minima or to prefer one over the other \footnote{ 
 Ofcourse the sign of higher derivatives might still be important for 
 consistency of the theory.}. 
 
 If we consider the case when the 'string' is stretched between two
 different critical points, $\phi_0 \neq \phi_1$, second 
 derivatives can be important. In particular, the sign of 
 $\phi_0^2 V^{(2)}(\phi_0) + \phi_1^2 V^{(2)}(\phi_1)$ should be negative to
 prevent the exponential from blowing up as $\l \rightarrow \infty$. 
 In other words, 
 perturbations around two minima will lead to trouble, perturbations 
 around a maximum and a minimum will be subject to constraints but 
 perturbations around two maxima will always be well defined in $(+,+)$ 
 theory. A similar analysis can be carried out for each of the theories. In 
 $(+,-)$ theory for example, we would find that perturbations around 
 a minimum-maximum configuration would always be safe. 

 As an alternative to the above argument, (atleast for the case 
 of a quadratic potential), we can study the worldsheet RG flow. Demanding 
 that this flow be free of singularities we reproduce exactly the same 
 results as above regarding 'safe' sectors of the theory. 
 We return to this point in the Appendix.   

\section{The Quadratic Approximation}

Let us assume that $T[X]$ is sufficiently well behaved to be
 expanded in a Taylor series   
\be
\label{potential}
T[X]= T_0 + T_{1i} X^i + \frac{1}{2} T_{2ij}X^i X^j +
 \frac{1}{3} T_{3ijk} X^i X^j X^k + ... ,
\ee
 where $X^i$ are spatial directions ($i=p+1,...,d-1$). 
 The general solution of equation of motion has the following form
\be
\label{exp}
 X^\mu(\tau,\sigma) = q^\mu + a_0^\mu \tau + b_0^\mu \sigma +
 \sum\limits_{n\neq 0} \frac{e^{-in\tau}}{n} ( i a_n^\mu \cos n\sigma
 + b_n^\mu \sin n\sigma) .
\ee
 For transverse directions $X^m$ ($m=0,1,...,p$) 
 we have the standard solution with Neumann conditions.
 For $X^i$ along the tachyon profile, we can 
 express $b_n^i$ as a function of $a_n^i$. In particular, for 
 the zero modes we have 
\be
\eta_{ij} b_0^i-2\pi\alpha'  \d_j T[q] =0,\,\,\,\,\,\,\,\,\,\,
\eta_{ij} b_0^i-2\pi\alpha' \d_j T[q+b_0 \pi]=0
\ee
 As $\l \rightarrow \infty$, the string ends become localized at extrema 
 of $T[X]$. There are no momentum modes $a^i_0$ since translational 
 symmetry is broken by the profile. 

No matter what the explicit form of the tachyonic field may be globally,  
a quadratic approximation is always reasonable locally around the extrema. 
Since these are precisely the regions we are interested in, it is sufficient 
for our purposes to consider a tachyonic field which is quadratic in $X^i$.
For the sake of simplicity we consider the profile to depend on just 
one direction. One can always choose a special basis such that $T[X]$ 
is brought to the form:
\be
\label{poquad}
 T[X] = \pm \frac{t_i}{2} (X^i-X_0^i) (X^i-X^i_1),
\ee
 where $t_{i}$ are (positive) dimensionful parameters. 
 The $\pm$ signs correspond to perturbations around minima and maxima 
 respectively. We want to study strings\footnote{
 Recall that conformal invariance is broken for 
 arbitrary $\l$, so a string theory interpretation is recovered 
 only at the fixed points of the RG flow.}
 that stretch between the D-branes which arise in the IR, at maxima 
 and minima of $T[X]$ .
 
We now solve for $X^i$, fixing the ends of the 'string' to lie at 
critical points of the quadratic field $T[X]$. 
In each given theory there arise four sectors, depending on whether  
 $\s = 0$ and $\pi$ lie at two minima, two maxima, a maximum and a minimum 
 or vice versa. The mode expansion for $X^i$ and the expressions for 
 the Virasoro generators are in general $\l$-dependent; the string 
 theory picture is recovered by taking the $\l \rightarrow \infty$ limit. 
 
 We start with a detailed analysis of the theory labelled $(-,-)$. 
 Boundary conditions are:
 \be
 \label{bc}
 X'^i - 2 \pi \alpha' \partial_i T[X]|_0 = 0 \;\;\;\; {\rm and} \;\;\;\;\;
 X'^i + 2 \pi \alpha' \partial_i T[X]|_{\pi} = 0.
 \ee
 Consider perturbations around a minimum and a maximum. 
 In this sector boundary conditions become: 
\be
\label{boundplm}
 X'^i - {\l}_i (X^i - x^i)|_0 = 0 \;\;\;\; {\rm and}\;\;\;\;\;
 X'^i - {\l}_i (X^i - \t{x}^i)|_{\pi} = 0
\ee
 where  $\lambda_i =2\pi\alpha' t_i$, $x^i=1/2(X_0^i +X^i_1)$ and 
 $\t{x}^i=1/2(\t{X}_0^i + \t{X}^i_1)$.
 and the mode expansion \footnote {The nontrivial $n$ dependence in 
 (\ref{solution}) comes from
 properly normalizing the eigenfunctions. This turns out to be crucial if we 
 are to avoid trouble when imposing canonical commutation relations.} is 
\be
\label{solution}
 X^i(\tau, \sigma) = q^i + b_0^i \sigma + \sqrt{2\alpha'} i
 \sum\limits_{n > 0} \frac{1}{\sqrt{n^2 + \l_i^2}}
 (\cos n\s + \frac{\lambda_i}{n} \sin n\s) [a^i_n e^{-in\tau}
 -  (a^i_n)^{+} e^{in\tau} ]
\ee
 where $n$ is an integer. In the UV, $b_0^i = 0$ and $q^i$ is left 
undetermined. However, for $\l_i \neq 0$ we find 
\be
\label{-+}
b_0^i = \frac{(\t{x}^i - x^i)}{\pi} \;\;\;\;\;\; {\rm and} \;\;\;\;\;\;
q^i = \frac{(\t{x}^i - x^i)}{\l_i \pi} + x^i.
\ee
 So as soon as $\l_i$ becomes non-zero, $b_0^i$ is fixed. However, 
 $q^i$ has an explicit $\l_i$ dependence, approaching $x^i$ as 
 $\l_i \rightarrow \infty$. It is customary to include a factor of $i$ 
 in the mode expansion for a 
 Neumann coordinate $X^i$, but not for a Dirichlet coordinate. Hence, 
 the way it stands, the above mode expansion is appropriate only for the 
 UV; we can apply it to the IR by absorbing the factor of $i$ in the 
 transformation $a^i_n \rightarrow ia^i_n$.
 Since this transformation is unitary, it preserves 
 commutation relations. 

 Canonical quantization, together with the
 identity\footnote{Since we are concerned with a Hermitian operator 
 $\d^2_\s$ on an interval $[0, \pi]$ and our boundary conditions are 
 of the required form, ours is a Sturm-Liouville problem. Properly normalized 
 eigenvectors of $\d^2_\s$ hence span an orthonormal basis in $L^2(0,\pi)$  
 and the expression (\ref{delta}) is simply the statement that this basis is 
 complete.} 
\be
\label{delta}
 \pi \delta(\s-\s') = 
 \sum\limits_{n\neq 0} \frac{n^2}{n^2+\lambda_i^2} (\cos n\s +
 \frac{\l_i}{n} \sin n\s) (\cos n\s' + \frac{\l_i}{n}\sin n\s'),
\ee 
results in the usual commutation relations between the modes 
\be
\label{commmode}
 [ a_n^i, (a_m^j)^+] = \eta^{ij} n \delta_{n-m}
 ,\,\,\,\,\,\,\,\,\,
 [ (a_n^i)^+, (a_m^j)^+] = [ a_n^i, a_m^j] = 0 .
\ee
 Hence the standard Fock space can be constructed for every
 value of $\l_i$.
 
 The transverse coordinates contribute to the Virasoro generators 
 in the usual way, but the $X^i$ contribution is modified so we 
 write it out explicitly:
 The zero mode is 
 \be
 L_0 = \bar{L}_0 = \frac{1}{4\alpha'} {b^i_0}^2 + \frac{1}{2} 
 \sum\limits_{n\neq 0} a^i_n a^i_{-n},
\ee
As usual for bosonic string theory, the Hamiltonian is given by 
 $H = L_0 -1$.
The Virasoro generators\footnote
{These expressions are appropriate for $\l_i \rightarrow \infty$. One 
can perform a unitary transformation to get the Virasoro generators 
which are appropriate for $\l_i \rightarrow 0$.}
are given by:
\be
\label{Lm}
 L_m = \frac{1}{\sqrt{2\alpha'}} \frac{\l_i-im}{\sqrt{m^2+\l_i^2}} a^i_m b^i_0
 -\frac{1}{2} \sum\limits_{n\neq m} \frac{( n +i\l_i) ((m-n)+ i\l_i)}
 {\sqrt{n^2+\l_i^2} \sqrt{(m-n)^2+\l_i^2}} a^i_n a^i_{m-n},
\ee
\be
\label{barLm}
 \bar{L}_m = -\frac{1}{\sqrt{2\alpha'}} \frac{\l_i+im}{\sqrt{m^2+\l_i^2}} 
 a^i_m b^i_0 - \frac{1}{2} 
 \sum\limits_{n\neq m} \frac{(n -i\l_i) ((m -n)- i\l_i)}{\sqrt{n^2+\l_i^2}
 \sqrt{(m-n)^2+\l_i^2}} a^i_n a^i_{m-n},
\ee
and obey the relevant algebras
\be
\label{VA} 
 [L_n, L_m] = (n-m) L_{n+m},\,\,\,\,\,\,\,\,\,\,\,\,\,\,
 [\bar{L}_n, \bar{L}_m] = (n-m) \bar{L}_{n+m}.
\ee
 However we see that $L_m \neq \bar{L}_m$ for $m\neq 0$,
 except at $\l_i =0$ and $\l_i =\infty$. 
 This is a manifestation of the now familiar statement that 
 Weyl invariance is broken for finite $\l_i$ and is restored only at 
the fixed points.

We now consider perturbations around a maximum and a minimum.  
Boundary conditions in this sector are the same as those in the 
minimum-maximum sector, with the only difference is that 
 $\l_i$ is replaced by $-\l_i$. Infact we find that all the results 
from our above discussion can be extended to this new sector of the theory, 
if we simply replace $\l_i$ by $-\l_i$ in the mode expansions 
 and Virasoro generators. $b_0^i$ is unaffected by this transformation 
 as it has no $\l_i$ dependence, but $q^i$ is now given by
 \be
 \label{+-}
 q^i = \frac{(x^i - \t{x}^i)}{\l_i \pi} + x^i.
 \ee
 Note that in the limit $\l_i \rightarrow \infty$, $q^i$ still 
 approaches $x^i$.

 As we remarked earlier, perturbations around maxima-minima, 
 or minima-maxima are related by a parity transformation. 
 The Virasoro generators have the symmetry
 \be
 \label{exter}
 \l_i \,\,\,\leftrightarrow \,\,\, -\l_i,\,\,\,\,\,\,\,\,\,\,\,
 \Leftrightarrow\,\,\,\,\,\,\,\,\,\,\, 
 L_m \,\,\,\leftrightarrow \,\,\, \bar{L}_m.
\ee 
 which can be interpreted as a flip in the orientation of the 'open string'
 due to a change in the sign of the boundary conditions . 
 
 Now look at perturbations around two minima. 
 Boundary conditions in this case are:
\be
\label{bcmmpp}
 X'^i - \l_i (X^i - \frac{1}{2}(X_0^i +X^i_1))|_{0} = 0,
\,\,\,\,\,\,\,\,\,\,\,\,\,\,
 X'^i + \l_i (X^i - \frac{1}{2}(X_0^i +X^i_1))|_{\pi} = 0. 
\ee
The mode expansions, Virasoro generators \footnote {Though we recover
 the standard Virasoro generators at fixed points $\l_i = 0$ and 
 $\l_i = \infty$, something slightly unusual happens for finite $\l_i$. 
 We can still define $L_n$ and $\bar{L_n}$ which obey the Virasoro 
 algebra, however there are now non-trivial space-time dependent phases 
 in the expressions for these generators.\cite{Husain:2001bt}} 
 and commutation relations are 
 given by the same formal expressions as before, the only difference being 
 that now $n, q^i$ and $b_0^i$ are functions of $\l_i$. In particular,
 $n$ is no longer an integer (except at the fixed points) and infact must 
 satisfy the condition
\be
\label{cond}
 \tan n\pi = \frac{2\l_i n}{n^2 - \l_i^2}.
\ee
 We find that $n(-\l_i)=-n(\l_i)$ in this parity invariant sector, so the 
 statement (\ref{exter}) does not hold here. 
 In the UV, $b_0^i = 0$ and $q^i$ is undetermined, whereas for $\l_i \neq 0$,
 we find
\be
\label{--}
q^i = \frac{\t{x}^i + x^i (1 + \l_i \pi)}{(2 + \l_i \pi)} \;\;\;\; {\rm and} 
\;\;\;\;\; 
b_0^i = \frac{\l_i(\t{x}^i - x^i)}{(2 + \l_i \pi)}
\ee
 Though the form of the $\l_i$ dependence in $q^i$ is different to 
 what it was in the two earlier cases, the asymptotic behaviour is the 
 same, $q^i \rightarrow x^i$ as $\l_i \rightarrow \infty$. 
 There is a drastic difference however, in $b_0^i$ which now 
 approaches $(x^i - \t{x}^i)/\pi$ in the IR. 
 Through this term, the Hamiltonian for a 'string' stretched between 
 two minima of $T[X]$ becomes $\lambda$ dependent. 

 The boundary conditions for the maxima-maxima sector are 
 related to those in the minima-minima sector by an 
 inversion in the sign of $\l_i$.
 In the UV we have the same result as before i.e $b_0^i = 0$ 
 and $q^i$ is undetermined. Along the flow however, we now have 
\be
\label{++}
 q^i (2-\l_i \pi) = x^i (1-\l_i \pi ) + \t{x}^i
 \;\;\;\;\; {\rm and} \;\;\;\;\;
 b_0^i(2-\l_i \pi)  = \l_i(x^i - \t{x}^i)
\ee
 which indicates a potential problem at $\l_i = 2/\pi$. Inorder for the 
 RG flow to be well defined, we must fix $x^i = \t{x}^i$, i.e, demand that
 $q^i = x^i$ and $b_0^i = 0$ for all $\l_i$. In 
 other words, we are not able to obtain strings that interpolate between 
 D-branes placed at two different maxima; the only consistent solutions that 
 exist are for a string that begins and ends on the same brane. 
 
  The same approach can be used to study the case 
 when $T[X]$ depends on more than one direction and we have 
 a set of couplings $(\l_{p+1}, ..., \l_{d-1})$. There are then 
 $2^{d-p}$ different sectors related to each other through world-sheet parity 
 $\Omega$. Also, the nature of the critical points will be much more 
 complicated.  
  
\section{Conclusions}

 In this paper, we look for consistent configurations of an open string 
 in a tachyonic background. Starting with a perfectly generic $T[X]$,
 we see that in the IR, D-branes should be localized either at zeros or 
 extrema of the tachyonic field, depending on the boundary conditions we 
 choose to impose.  We then study the explicit case when $T[X]$ is 
 quadratic, as this is an exactly solvable model and is also a 
 good approximation around the regions where D-branes will appear. 
 Though the boundary conditions in the IR force us to localize D-branes at 
 critical points of $T[X]$, no distinction is made between maxim and minima
 so we consider both. However, proceeding along these lines, we 
 find that there is a discrepancy when we put the picture together.
  
 As we flow to the IR limit in $(-,-)$ theory, we find strings which 
 stretch between maxima and minima, or two different minima but 
 none that interpolate between two different maxima. 
 Perhaps this is a signal that we allow D-branes to appear only at the 
 minima of the tachyonic field since all possible strings which can exist 
 in such a configuration are allowed, and the theory thus truncated 
 becomes completely consistent. This coincides exactly with the picture 
 proposed by Harvey et al\cite{Harvey:2000na} and supported by the g-theorem.

 The discussion in the above paragraph carries over in totality to the 
 $(+,+)$ theory, the only difference being that here there are no strings 
 interpolating between two minima since the theories are equivalent
 under interchange of maxima and minima. 

 \appendix
 \section{Appendix}
 A quadratic approximation for $T[X]$ is, as we have argued, reasonable around
 critical points of the tachyonic field. However, the analysis in Section 3
 does not fully exploit the possibilities of this approximation. We have 
 restricted ourselves to maxima and minima which are identical in shape 
 and size and are related simply by an inversion. This leads to the fact that
 second derivatives at all critical points have the same magnitude and 
 differ only in sign. This would be the case for a $\sin$ or $\cos$ 
 potential ofcourse, but in general one could picture a more irregular 
 form for $T[X]$ where the shapes of maxima and minima vary and consequently 
 the second derivatives at the critical points may differ in their 
 absolute values as well. We now analyse this situation, using the methods 
 already outlined in Section 3. 
 
 Consider for example $(-,-)$ theory. Around the extrema where string ends 
are confined, $T[X]$ can be locally approximated
 by  
 \be
 T[X] = \frac{t_i}{2}  A (X^i - X^i_0)(X^i - X^i_1) \;\; (\s = 0) \;\;\; 
 {\rm and} \;\;\;
 T[X]|_{\pi} = \frac{t_i}{2} B (X^i - \t{X}^i_0)(X^i - \t{X}^i_1) \;\; (\s = \pi)
 \ee 
 where A and B can be positive or negative corresponding to minima and 
 maxima respectively. 
 The boundary conditions become 
 \be
 \label{bcc}
 X'^i - \l A (X^i - x^i) = 0 \; {\rm at} \; \s = 0 
 \;\;\;\; {\rm and} \;\;\;\;\;
 X'^i + \l B (X^i - \t{x}^i) = 0 \; {\rm at} \; \s = \pi.
 \ee
 So $q^i$ and $b_0^i$ are given by
 \be
 q^i[ A + B + \l A B \pi] = (\l A B \pi + A)x^i + b \t{x}^i 
 \ee
 and
 \be
 b^i_0[ A + B + \l A B \pi] = \l A B (\t{x}^i - x^i). 
 \ee
 Hence there is a possible singularity in the flow when 
 $[ A + B + \l A B \pi] =0$, unless we restrict both
 string ends to the same critical point. We do not however want to 
 apply this restriction in general, since our aim is to 
 obtain strings that stretch between two different D-branes in the IR.  
 Keeping in mind that $\l$ must always be positive, we see that the only 
 case where there will never be a singularity, irrespective of the values 
of A and B, is when they are both positive. 
 
 Hence, in general for an irregularly shaped tachyonic background field, the 
 only sector in which the renormalization group flow never develops a 
 singularity is that which corresponds to perturbations around two minima. 
 This extends the analysis of section 3 and strengthens the conclusions 
 drawn in section 4. A similar analysis can be performed for the 
 three remaining theories. In each case we find that there is just 
 one sector which is always free from singularities; ofcourse this sector 
 differs from one theory to another. In each case though, it coincides 
 exactly with the 'safe' sector that results from the path integral 
 argument towards the end of Section 2. 
 
\begin{flushleft} {\Large\bf Acknowledgments} \end{flushleft}

We thank Paolo di Vecchia for organizing a wonderful 
meeting, and for inviting us to present this talk in a stimulating 
atmosphere. We would also like to acknowledge enlightening and encouraging 
discussions with Alberto Lerda following our presentation. During the course
of this work, we had many useful conversations with Ingemar Bengtsson, 
Ansar Fayyazuddin, Hans Hansson and Ulf Lindstr\"om for which we are 
very grateful. 

\noindent


\end{document}